\documentclass[onecolumn,12pt]{aastex6}
\usepackage{lineno,setspace}

\shorttitle{Upper Limits on O$_2$ in Pluto's Atmosphere}
\shortauthors{Kammer et~al.}

\begin{document}

\title{\textit{New Horizons} Upper Limits on O$_2$ in Pluto's Present Day Atmosphere}

\author{J.~A.~Kammer}
\author{G.~R.~Gladstone}
\affil{Southwest Research Institute\\San Antonio, TX 78238, USA}
\email{jkammer@swri.edu}

\author{S.~A.~Stern}
\author{L.~A.~Young}
\author{C.~B.~Olkin}
\author{A.~Steffl}
\affil{Southwest Research Institute\\Boulder, CO 80302, USA}

\author{H.~A.~Weaver}
\affil{Johns Hopkins Applied Physics Laboratory\\Laurel, MD 20723, USA}

\author{K.~Ennico}
\affil{NASA Ames Research Center\\Moffett Field, CA 94035, USA}

\author{and the \textit{New Horizons} Atmospheres and Alice UV Spectrograph Teams}

\begin{abstract}
The surprising discovery by the \textit{Rosetta} spacecraft of molecular oxygen (O$_2$) in the coma of comet~67P/Churyumov-Gerasimenko \citep{bieler2015} challenged our understanding of the inventory of this volatile species on and inside bodies from the Kuiper Belt. That discovery motivated our search for oxygen in the atmosphere of Kuiper Belt planet Pluto, because O$_2$ is volatile even at Pluto's surface temperatures. During the \textit{New Horizons} flyby of Pluto in July 2015, the spacecraft probed the composition of Pluto's atmosphere using a variety of observations, including an ultraviolet solar occultation observed by the Alice UV spectrograph \citep{stern2015,gladstone2016,young2017}. As described in these reports, absorption by molecular species in Pluto's atmosphere yielded detections of N$_2$, as well as hydrocarbon species such as CH$_4$, C$_2$H$_2$, C$_2$H$_4$, and C$_2$H$_6$. Our work here further examines this data to search for UV absorption from molecular oxygen (O$_2$), which has a significant cross section in the Alice spectrograph bandpass. We find no evidence for O$_2$ absorption, and place an upper limit on the total amount of O$_2$ in Pluto's atmosphere as a function of tangent height up to 700 km. In most of the atmosphere this upper limit in line of sight abundance units is $\sim$3$\times$10$^{15}$ cm$^{-2}$, which depending on tangent height corresponds to a mixing ratio of 10$^{-6}$ to 10$^{-4}$, far lower than in comet 67P/CG.

\end{abstract}


\section{Introduction}

\textit{In situ} mass spectroscopy measurements aboard the \textit{Rosetta} spacecraft at comet~67P/Churyumov-Gerasimenko (67P/CG) detected molecular oxygen (O$_2$), which represents the first detection of O$_2$ in a cometary coma \citep{bieler2015}. Subsequently, FUV stellar occultation observations \citep{keeney2016} by the \textit{Rosetta} Alice ultraviolet spectrograph \citep{stern2007} also detected O$_2$. The molecular oxygen in comet~67P/CG was found to be present at surprisingly high levels (several to several tens of percent relative to the line of sight column abundance of H$_2$O), suggesting that the O$_2$ is primordial, rather than the byproduct of coma chemistry, and therefore has likely been stored in the nucleus of the comet since its formation. O$_2$ has been found in atmospheres throughout the solar system, from terrestrial planets like Earth to icy moons such as Europa \citep{hall1995}, and can also be found produced in the rings of Saturn \citep{johnson2006}. However, comet~67P/CG originated in the distant reaches of the Kuiper Belt. Because such comets are theorized candidates for the building blocks of bodies like Pluto, this naturally motivates a search for molecular oxygen in the atmosphere of Pluto by making use of the datasets acquired by the \textit{New Horizons} mission during its flyby in July 2015.

One of the powerful tools for atmospheric study on the \textit{New Horizons} spacecraft is the Alice extreme-/far-ultraviolet imaging spectrograph \citep[for a thorough instrument description, see][]{stern2008}. This instrument has a bandpass of 52 to 187 nm, a wavelength region that is highly diagnostic for signatures of absorption by molecular species. Initial results from the analysis of a UV solar occultation of Pluto observed by this instrument yielded strong detections of both N$_2$ and CH$_4$, detections of minor hydrocarbon species such as C$_2$H$_2$, C$_2$H$_4$, and C$_2$H$_6$, as well as a measurement of the absorption by the haze that enshrouds the planet \citep{stern2015,gladstone2016,young2017}. Because molecular oxygen also has significant absorption in this region of the UV \citep{ackerman1970,brion1979,gibson1980}, this work delves further into the \textit{New Horizons} Alice occultation dataset with the goal of detecting or constraining the abundance of O$_2$ in the atmosphere of Pluto. In \S\ref{o2inUV}, we compare the absorption of O$_2$ to other species detected in the UV solar occultation, and set upper limits on O$_2$ line of sight column abundances as a function of tangent height. In \S\ref{comptoCG}, we discuss these results at Pluto within the context of \textit{Rosetta} findings at comet~67P/CG. Finally, in \S\ref{conclude}, we consider possible explanations for the dramatic difference we observe in O$_2$ abundances, and suggest future observations that could improve our understanding of the prevalence of O$_2$ in the Kuiper Belt.

\section{O$_2$ in the Ultraviolet}\label{o2inUV}

\subsection{Comparison to Other UV Absorbing Species}

\begin{figure}[htb!]
\epsscale{1.0}
\plotone{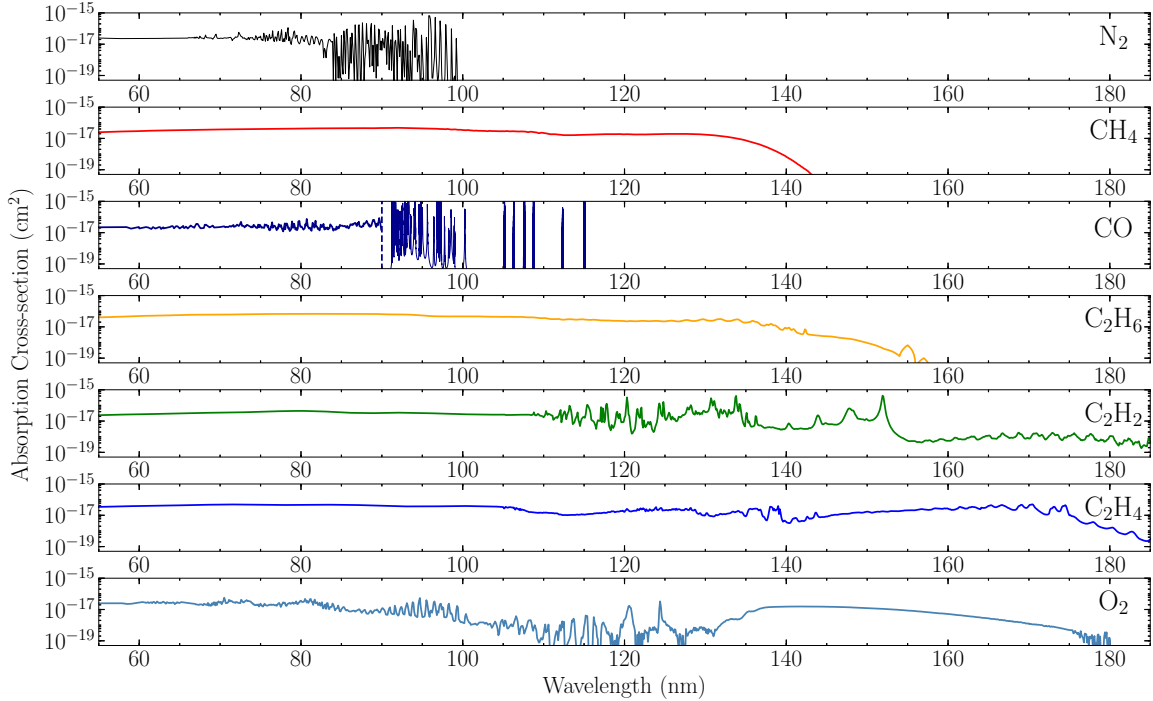}
\caption{Absorption cross-section of O$_2$ in the ultraviolet compared to those of several other species found in the atmosphere of Pluto. Note that CO is represented by both low and high resolution cross-sections where available, with a transition region from low to high located at 90 nm; the other species have relatively lower resolution cross sections at the relevant temperatures. The sources for the cross-sections used in this work are compiled in Table~\ref{crosstab}.}
\label{crossfig}
\end{figure}

Figure~\ref{crossfig} compares the general features of the UV cross-section of O$_2$ with other species detected in Pluto's atmosphere, including N$_2$, CO, and CH$_4$ along with its photochemical derivatives (C$_2$H$_2$, C$_2$H$_4$, C$_2$H$_6$). Disentangling the combined effects of absorption by these various species can be achieved due to their distinctly different cross sections as a function of wavelength. References for these cross-sections are listed in Table~\ref{crosstab}.

\begin{deluxetable*}{ccl}
\tabletypesize{\scriptsize}
\tablecaption{Absorption Cross-Sections of Relevant Species}
\tablecolumns{3}
\tablewidth{0pt}
\tablehead{
\colhead{Species} & \colhead{Name} & \colhead{Source}}
\startdata
N$_2$ & Nitrogen & \cite{shaw1992}; \cite{heays2011}\\
CH$_4$ & Methane & \cite{lee2001}; \cite{kameta2002}; \cite{chen2004}\\
CO & Carbon Monoxide & \cite{chan1993}; \cite{visser2009}; \cite{stark2014}\\
C$_2$H$_6$ & Ethane & \cite{kameta1996}; \cite{lee2001}; \cite{chen2004}\\
C$_2$H$_2$ & Acetylene & \cite{nakayama1964}; \cite{cooper1995}; \cite{wu2001}\\
C$_2$H$_4$ & Ethylene & \cite{cooper1995}; \cite{wu2004}\\
O$_2$ & Oxygen & \cite{ackerman1970}; \cite{brion1979}; \cite{gibson1980}
\enddata
\label{crosstab}
\end{deluxetable*}

\begin{figure}[htb!]
\epsscale{1.0}
\plotone{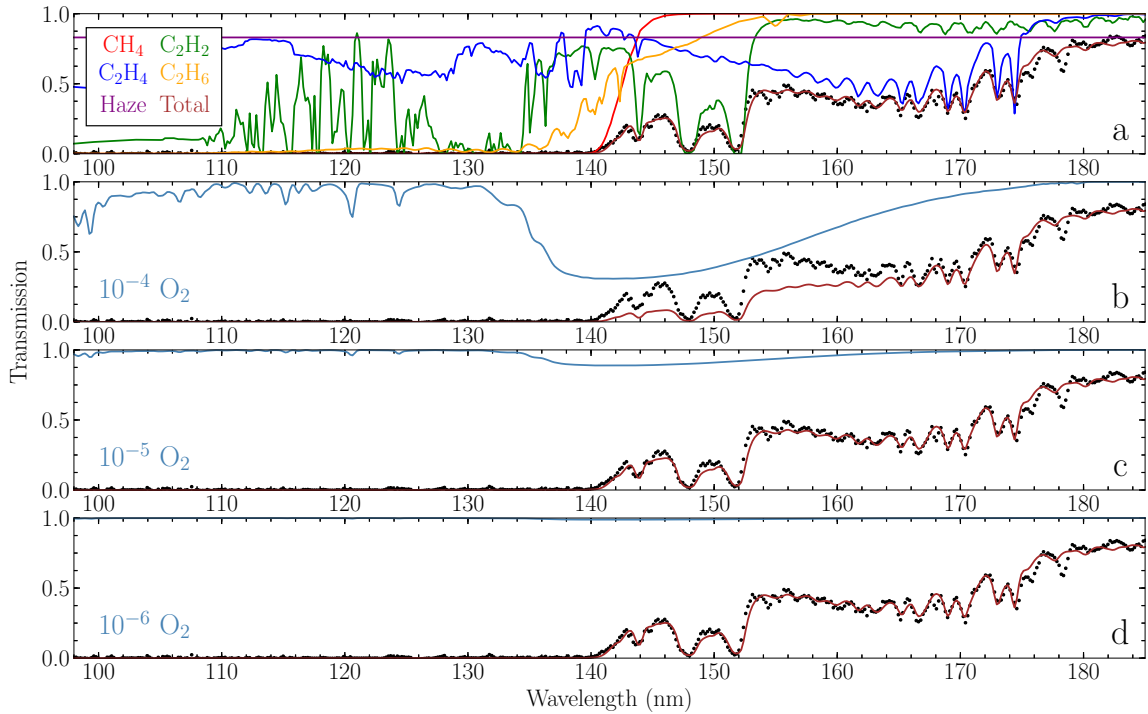}
\caption{Effects of various assumed O$_2$ mixing ratios on the observed ingress Pluto solar occultation spectrum at 200 km tangent height above the surface compared to the observed spectrum at this altitude (black points). Panel a: Best fit contributions to absorption by each of the detected hydrocarbon species from Y17, and the total model transmission (brown). Panels b-d: Absorption by O$_2$ is added to the existing best fit model in bulk atmospheric mixing ratios of 10$^{-4}$, 10$^{-5}$, and 10$^{-6}$, respectively. Only O$_2$ mixing ratios at or below 10$^{-5}$ remain consistent with the data at this tangent height.}
\label{specfig}
\end{figure}

\begin{figure}[htb!]
\epsscale{1.0}
\plotone{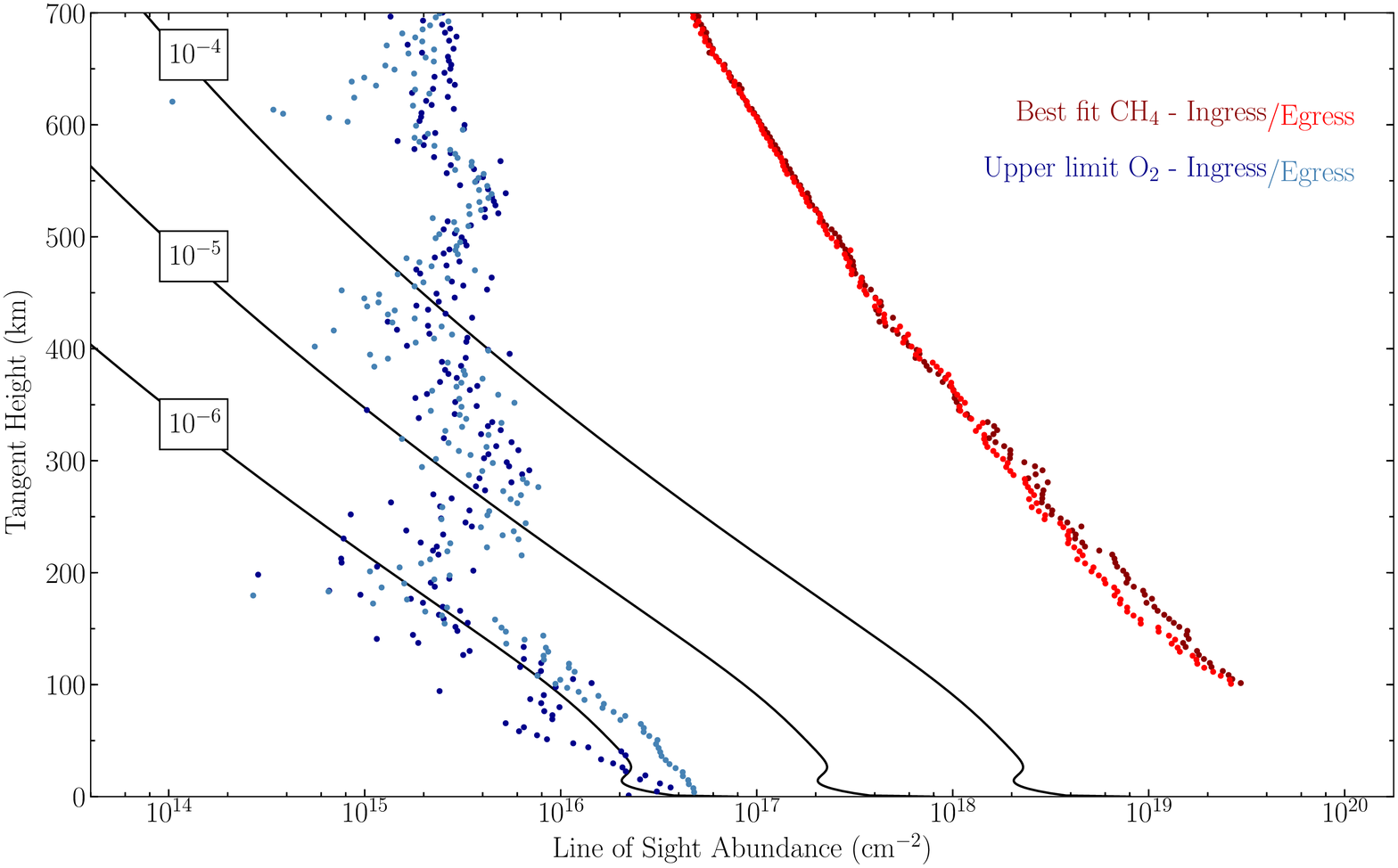}
\caption{Profiles of the derived upper limit on the line of sight column abundances of O$_2$ (blue, light blue) during the Pluto solar occultation ingress (entry) and egress (exit) at tangent heights up to 700 km above the surface. Solid black lines indicate line of sight abundances equivalent to constant mixing ratios of 10$^{-4}$, 10$^{-5}$, and 10$^{-6}$ relative to the N$_2$ model of Y17. The best fit values for CH$_4$ line of sight abundances (red, light red) from the analysis of Y17 are also shown for comparison.}
\label{profilefig}
\end{figure}

\cite{young2017} (hereafter Y17) derived line of sight abundances of N$_2$, CH$_4$, C$_2$H$_2$, C$_2$H$_4$, C$_2$H$_6$, and a neutral-absorbing haze to match the observed Pluto occultation spectra at wavelengths from 55-65 nm and 100-185 nm. The results of Y17 provide our initial best fit models for wavelength-dependent transmission at each level in the atmosphere. We note that N$_2$, the hydrocarbons, and haze together account for the bulk of all atmospheric absorption detected during the Alice ultraviolet solar occultation. Here we reanalyze this same dataset at the longer wavelength end in order to determine how much O$_2$ can be accommodated in a modeled atmosphere given the Alice detection limits.

As demonstrated in the top panel of Figure~\ref{specfig}, which shows model results from Y17, followed in lower panels by modeled absorption due to O$_2$ fixed at 200 km altitude, we found that sensitivity to O$_2$ remains high despite the absorptions of other species. We tested three simple cases in which the mixing ratio of O$_2$ was fixed to a constant value with altitude, then calculated the effect on the atmospheric absorption in the UV, while keeping all other species abundances the same. This yielded an initial estimate of the maximum tolerable mixing ratio of O$_2$. More specifically, when using this approach the Pluto solar occultation is sensitive to and clearly rules out absorption by O$_2$ at the $\sim$10$^{-5}$ mixing ratio level or higher.

\subsection{Vertical Profiles of Upper Limits on O$_2$ Abundance}

We note, however, that constraints on O$_2$ abundances using this dataset depend upon the tangent height of the observed spectra. Because a primary physical quantity measured by the occultation is the absorbing species line of sight (LOS) column abundance (cm$^{-2}$), we also report the O$_2$ LOS abundance upper limits as a function of tangent height for both ingress and egress. During this part of our analysis, the contribution to UV absorption from O$_2$ was added as another free parameter in a simultaneous fit to the spectra, as done in Y17. The upper limit on LOS abundance of O$_2$ was then derived at each tangent height based on an optimized model spectral fit, briefly summarized here (see Y17 for a full description and discussion of the procedure). Starting at a tangent height of 2000 km altitude, spectra from each level in the atmosphere were iteratively fit down toward the surface. To retrieve the parametrized line-of-sight abundances from each spectrum, the spectrum from the previous level acted as an initial guess in a weighted Levenberg-Marquardt least-squares fit to minimize the weighted sum of squared residuals, or $\chi ^2$. The upper limits derived for LOS abundances of O$_2$ are shown in Figure~\ref{profilefig}.

In most of the upper atmosphere, the limit on the O$_2$ LOS column abundance is $\sim$3$\times$10$^{15}$ cm$^{-2}$. Depending on tangent height, this LOS abundance is generally equivalent to a constant mixing ratio model of 10$^{-4}$ at 450 km, 10$^{-5}$ at 250 km, and 10$^{-6}$ at 150 km, becoming smaller as the tangent height decreases. This trend significantly weakens below 150 km, because the transmission of sunlight begins to approach zero at all wavelengths that are sensitive to O$_2$ absorption, and thus additional O$_2$ absorption no longer has a significant effect on the model fit.

\section{Pluto Compared to Comet 67P/CG}\label{comptoCG}

While the \textit{in situ} results from the \textit{Rosetta} mission at comet~67P/CG reported an O$_2$/H$_2$O ratio, it is far more useful at Pluto to compare the measured O$_2$/CO and O$_2$/N$_2$ ratios, because the temperatures in Pluto's atmosphere are too cold for H$_2$O to remain in gas phase but are not too cold for CO or N$_2$ to remain in gas phase. The relative abundances of these species in the atmosphere of Pluto and the coma of comet~67P/CG may then shed light on the provenance of the O$_2$.

For comet~67P/CG, the O$_2$/CO ratio varied from about 0.1 to 1.0 \citep{bieler2015}. At Pluto, the mixing ratio of CO has been most recently estimated from observations by ALMA to be 515$\pm$40 ppm \citep{lellouch2017}. Taken together with the upper limit from this work of $\sim$10$^{-5}$ O$_2$ mixing ratio in Pluto's entire atmospheric column, this implies a Pluto O$_2$/CO upper limit of $\sim$2$\times$10$^{-2}$. This is 5-50 times lower than the O$_2$/CO ratio found in the coma of comet~67P/CG. 

In the case of O$_2$/N$_2$ ratios, the difference is even more extreme. For comet~67P/CG, the O$_2$/N$_2$ ratio varied from about 20-100, with much more O$_2$ detected in outgassing from the comet than N$_2$. This is in stark contrast to Pluto's atmosphere, which is dominated by N$_2$, and which has an upper limit O$_2$ mixing ratio relative to N$_2$ of $\sim$10$^{-5}$. 

\section{Conclusions}\label{conclude}

We suggest several explanations for the relative lack of atmospheric O$_2$ on Pluto, which indicate that this warrants further study. First, it may be that comet~67P/CG is simply not typical of the material that went into the formation of Pluto. Alternatively, perhaps Pluto contains a similar inventory of primordial O$_2$, but it is not found in appreciable amounts in the present day atmosphere, i.e., it may be sequestered in the planet's interior. Or as another alternative, O$_2$ inside Pluto may have been chemically converted into other oxygen-bearing species such as H$_2$O, CO, CO$_2$, or minerals, owing to the expected significantly different thermal and chemical evolution of Pluto, which is $\sim$10$^{9}$ times as massive as comet~67P/CG. 

In order to make further progress, we suggest that new observations are needed. Specifically, it is necessary to determine how typical the O$_2$ abundance in comet~67P/CG is for Kuiper Belt comets as a whole; \cite{keeney2016} have shown that the O$_2$ in comet~67P/CG can also be detected by FUV stellar occultation techniques, opening the possibility that future observations with the Hubble Space Telescope or similarly capable future observatories could address this issue.

\acknowledgements{
We thank the entire \textit{New Horizons} team for making the exploration of Pluto possible, and we thank NASA for financial support of the \textit{New Horizons} project.
}

\end{document}